\documentclass[twocolumn,showpacs,preprintnumbers,amsmath,amssymb]{revtex4}

\usepackage{graphicx}% Include figure files
\usepackage{dcolumn}% Align table columns on decimal point
\usepackage{bm}% bold math

\begin{document}
\draft

\title{Accurate numerical solution to the finite-size Dicke model}
\author{Qing-Hu Chen$^{1,2}$, Yu-Yu Zhang$^{2}$, Tao Liu$^{3}$, and Ke-Lin Wang
$^{3,4}$}

\address{
$^{1}$ Center for Statistical and Theoretical Condensed Matter
Physics, Zhejiang Normal University, Jinhua 321004, P. R. China  \\
$^2$ Department of Physics, Zhejiang University, Hangzhou 310027,
P. R. China \\
$^{3}$Department of Physics, Southwest University of  Science and Technology, Mianyang 621010, P.  R.  China\\
$^{4}$Department of Modern Physics, University of  Science and
Technology of China,  Hefei 230026, P.  R.  China
 }
\date{\today}

\begin{abstract}
By using extended bosonic coherent states,  a new  technique to
solve the Dicke model exactly is proposed in the numerical sense.
The accessible system size is two orders of magnitude higher than
that reported in literature. Finite-size scaling for several
observables, such as the ground-state energy, Berry phase, and
concurrence are analyzed.  The existing discrepancy for the
scaling exponent of the concurrence is reconciled.
\end{abstract}


\maketitle

The Dicke model\cite{dicke} describes the interaction of N two-level
atoms (qubits) with a single bosonic mode and has been paradigmatic
example of collective quantum behavior. It exhibits a
"superrandiant" quantum phase transition (QPT)\cite{Sachdev} in the
thermodynamic limit, which was first studied by Hepp and Lieb
\cite{lieb} at weak coupling. In recent years, the Dicke model has
attracted considerable attentions due to the progress in the QPT.
First, entanglement, one of the most striking consequences of
quantum correlation in many-body systems, shows a deep relation with
the QPT\cite {oster}. Understanding the entanglement is also a
central goal of quantum information science. Second, Berry phase has
been recently drawn in the study of the QPT \cite{Carollo}. A
drastic change at critical point in the QPT may be reflected in the
geometry of the Hilbert space, the geometric phase may capture the
singularity, and therefore signal the presence of the QPT.

In the thermodynamic limit, the  Dicke model is exactly soluble in
the whole coupling range, based upon the Holstein-Primakoff
transformation of the angular momentum algebra\cite{Emary}. For
finite N, the Dicke model is in general nonintegrable. The
finite-size correction in this system has been shown to be crucial
in the understanding the properties of the entanglement \cite
{Lambert1,Buzek,Lambert2,liberti,vidal,reslen} and the Berry
phase\cite {plastina}, in which one can characterize universality
around the critical point in the QPT. However, a convincing exact
treatment of the finite-size Dicke model is still lacking. To the
best of our knowledge, the finite-size studies
are limited to numerical diagonalization in Bosonic Fock state \cite{Emary}%
\cite{Lambert1}\cite{Lambert2} in small size system $N\leq 35$, the
adiabatic approximation \cite{liberti}, and $1/N$ expansion based on
modified Holstein-Primakoff approach\cite{vidal}.

Recently, the Dicke model is closely related to many fields in
quantum optics and condensed matter physics, such as the
superradiant behavior by an ensemble of quantum dots
\cite{Scheibner} and Bose-Einstein condensates \cite {Schneble}, and
coupled arrays of optical cavities used to simulate and study the
behavior of strongly correlated systems\cite{Hartmann}. The
finite-size Dicke model itself is also of practical interest. It was
observed that the Dicke model for finite N can be realized in
several solid-state systems. One Josephson charge qubit coupling to
an electromagnetic resonator \cite {Wallraff} can be described by
$N=1$ Dicke model , which is just special case of spin-1/2 (Rabi
Hamiltonian). The features of the superconducting quantum
interference device coupled with a nanomechanical resonator may be
captured in the framework of the finite-size Dicke model
\cite{squid}.

In this paper, we propose an exact technique to solve Dicke model
numerically for finite N by means of extended bosonic coherent
states. The correlations among bosons are added step by step until
further corrections will not change the results. Some important
quantities are calculated exactly.

Without the rotating-wave approximation, the Hamiltonian of $N$ identical
qubits interacting with a single bosonic mode is originally given by

\begin{equation}
H=\omega a^{+}a+\Delta J_z+\frac{2\lambda }{\sqrt{N}}(a^{+}+a)J_x,
\label{hamiltonian}
\end{equation}
where $a^{+}$ and $a$ are the field annihilation and creation
operators, $  \Delta $ and $\omega $ are the transition frequency of
the qubit and the frequency of the single bosonic mode, $\lambda$ is
the coupling constant. $  J_x$ and $J_z$ are the usual angular
momentum. There is a conserved parity operator $\Pi =e^{i\pi
(Jz+N/2+a^{+}a)}$, which commutes with the Hamiltonian (1). For
convenience, we use a transformed Hamiltonian with a rotation around
an $y$ axis by an angle $\frac \pi 2$ ,
\begin{equation}
H=\omega a^{+}a-\frac \Delta 2(J_{+}+J_{-})+\frac{2\lambda }{\sqrt{N}}%
(a^{+}+a)J_z,,  \label{hamiltonian1}
\end{equation}
where $J_{+}$ and $J_{-}$ are the angular raising and lowing operators and
obey the SU(2) Lie algebra $[J_{+},J_{-}]=2J_z,[J_z,J_{\pm }]=\pm J_{\pm }$.
So the Hilbert space of this algebra is spanned by the Dicke state $\{\left|
j,m\right\rangle ,m=-j,-j+1,...j-1,j\}$ with $j=N/2$, which is the
eigenstate of $J^2$ and $J_z$ with the eigenvalues $j(j+1)$ and $m$.

The Hilbert space of the total system can be expressed in terms of
the basis $\{\left| \varphi _m\right\rangle _b\bigotimes \left|
j,m\right\rangle \}$, where only the state of bosons$\left| \varphi
_m\right\rangle _b$ is to be determined. A "natural" basis for
bosons is Fock state $\left| l\right\rangle
_b=[(a^{+})^l/\sqrt{l!}]\left| 0\right\rangle _b$. In the Dicke
model, the bosonic number is not conserved, so the bosonic Fock
space has infinite
dimensions, the standard diagonalization procedure (see, for example, Ref. [%
\cite{Emary}\cite{Lambert1}\cite{Lambert2}]) is to apply a
truncation procedure considering only a truncated number of bosons.
Typically, the convergence is assumed to be achieved if the
ground-state energy is determined within a very small relative
errors. Within this method, one has to diagonalize very large,
sparse Hamiltonian in strong coupling regime and/or in adiabatic
regime. Furthermore, the calculation becomes prohibitive for larger
system size since the convergence of the ground-state energy is very
slow. Interestingly, this problem can be circumvented in the
following procedure.

By the displacement transformation $A_m=a+g_m$ with $g_m=2\lambda m/\omega
\sqrt{N}$, the Schr$\stackrel{..}{o}$ dinger equation can be described in
columnar matrix, and its $m$ row reads
\begin{eqnarray}
&&-\Delta j_m^{+}\left| \varphi _m\right\rangle _b\left| j,m+1\right\rangle
-\Delta j_m^{-}\left| \varphi _m\right\rangle _b\left| j,m-1\right\rangle
\nonumber \\
&&\omega (A_m^{+}A_m-g_m^2)\left| \varphi _m\right\rangle _b\left|
j,m\right\rangle =E\left| \varphi _m\right\rangle _b\left| j,m\right\rangle,
\end{eqnarray}
where $j_m^{\pm }=\frac 12\sqrt{(j(j+1)-m(m\pm 1)}$. Left multiplying $%
\{\left\langle n,j\right| \}$ gives a set of equations
\begin{eqnarray}
&&-\Delta j_n^{+}\left| \varphi _{n+1}\right\rangle _b-\Delta j_n^{-}\left|
\varphi _{n-1}\right\rangle _b+\omega (A_n^{+}A_n-g_n^2)\left| \varphi
_n\right\rangle _b  \nonumber \\
&=&E\varphi _n\left| \varphi _n\right\rangle _b,
\end{eqnarray}
where $n=-j,-j+1,...j$.

Note that the linear term for the bosonic operator $a(a^{+})$ is removed,
and a new free bosonic field with operator $A(A^{+})$ appears. In the next
step, we naturally choose the basis in terms of this new operator, instead
of $a(a^{+})$, by which the bosonic state can be expanded as
\begin{eqnarray}
\left| \varphi _n\right\rangle _b
&=&\sum_{k=0}^{N_{tr}}c_{n,k}(A_n^{+})^k\left| 0\right\rangle _{A_n}
\nonumber \\
&=&\sum_{k=0}^{N_{tr}}c_{n,k}\frac 1{\sqrt{k!}%
}(a^{+}+g_n)^ke^{-g_na^{+}-g_n^2/2}\left| 0\right\rangle _a,
\label{wavefunction}
\end{eqnarray}
where $N_{tr}$ is the truncated bosonic number in the Fock space of $A(A^{+})
$. As we know that the vacuum state $\left| 0\right\rangle _{A_n}$ is just a
bosonic coherent-state in $a(a^{+})$ with an eigenvalue $g_n$\cite
{chen,qin,liu}. So this new basis is overcomplete, and actually does not
involve any truncation in the Fock space of $a(a^{+})$, which highlights the
present approach. It is also clear that many-body correlations for bosons
are essentially included in extended coherent states (5). Left multiplying
state $_{A_n}\left\langle l\right| $ yields
\begin{eqnarray}
&&\omega (l-g_n^2)c_{n,l}-\Delta
j_n^{+}\sum_{k=0}^{N_{tr}}c_{n+1,k}\text{ }
_{A_n}\langle l\left| k\right\rangle _{A_{n+1}}  \nonumber \\
- &&\Delta j_n^{-}\sum_{k=0}^{N_{tr}}c_{n-1,k}\text{ }_{A_n}\langle l\left|
k\right\rangle _{A_{n-1}})=Ec_{n,l},
\end{eqnarray}
where
\begin{eqnarray}
_{A_n}\langle l\left| k\right\rangle _{A_{n-1}} &=&(-1)^lD_{l.k},  \nonumber
\\
_{A_n}\langle l\left| k\right\rangle _{A_{n+1}} &=&(-1)^kD_{l.k},
\end{eqnarray}
with
\[
D_{l,k}=e^{-G^2/2}\sum_{r=0}^{min[l,k]}\frac{(-1)^{-r}\sqrt{l!k!}G^{l+k-2r}}{
(l-r)!(k-r)!r!},G=\frac{2\lambda }{\omega \sqrt{N}}
\]
. Eq. (6) is just a eigenvalue problem, which can be solved by the exact
Lanczos diagonalization approach in dimensions $(N+1)(N_{tr}+1)$\cite{tian}.
The integral of motion due to the parity $\Pi $ can simplify the computation
further. To obtain the true exact results, in principle, the truncated
number $N_{tr}$ should be taken to infinity. Fortunately, it is not
necessary. It is found that finite terms in state (5) are sufficient to give
very accurate results with a relative errors less than $10^{-6}$, in the
whole coupling range. We believe that we have exactly solved this model
numerically.

\begin{figure}[tbp]
\centering
\includegraphics[width=8cm]{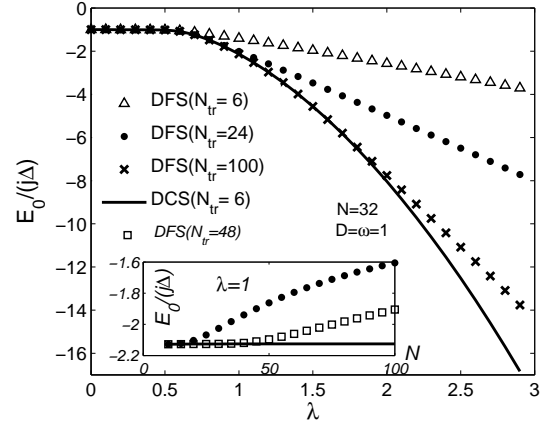}
\caption{The scaled ground-state energy $E_{0}/(j\Delta )$ as a
function of $\lambda $ by DCS and DFS with different  truncated
numbers. The inset show the scaled ground-state energy as a function
of the $N$ for fixed $ \lambda=1 $. } \label{fig1}
\end{figure}

To show the effectiveness of the present approach, we first
calculate the ground-state energy $E_0$ by diagonalization in
coherent-states (DCS) as described above , and compare with those by
numerical diagonalization in Fock state (DFS) \cite{Emary} for N=32.
For convenience, we introduce two dimensionless parameters $D=\Delta
/\omega $ and $\alpha =4\lambda ^2/\Delta \omega$. Fig. 1 shows the
scaled ground-state energy $E_{0}/(j\Delta )$ as a function of
$\lambda $ when the Hamiltonain is on a scaled resonance $\Delta
=\omega=1 $ . For the same truncated number $N_{tr}=6$, the
ground-state energy by DCS is much lower than that by DFS. With the
increase of $N_{tr}$, DFS results approach the DCS one monotonously
in the stronger coupling regime. One can find that the ground-state
energy by DCS with $N_{tr}=6$ is even much better than that by DFS
with $ N_{tr}=100$. In addition, the DFS results become worse and
worse as the N increases for fixed $N_{tr}$ and $\lambda $, as
clearly shown in the inset of Fig. 1.

\begin{figure}[tbp]
\centering
\includegraphics[width=8cm]{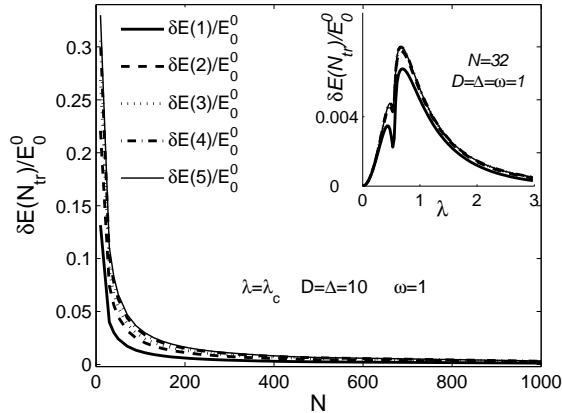}
\caption{The relative ground-state energy difference as a function
of $N$ for different truncated numbers $N_{tr}$ at critical point.
The inset show $ \delta E(N_{tr})$ as a function of $\lambda$ at
fixed $N$ for different $N_{tr}$. } \label{fig2}
\end{figure}

The ground-state  energy is non-analytical at the critical point of
the QPT in the thermodynamic limit. To show the effect of the
precursor of the QPT
on the present approach, we display the relative ground-state energy difference $%
\delta E(N_{tr})/E_{0}^0=(E_{0}^{N_{tr}}-E_{0}^0)/E_{0}^0$ as
function of $N$ at the critical point $\lambda_c$ in Fig. 2. It is
interesting to find that as $N$ increases, the exact results can be
obtained within smaller truncated number. This feature facilitates
the calculation for rather large system size. In the inset of the
Fig. 2, we can find that in both the weak and strong coupling
regime, the results converge rapidly with the truncated number
$N_{tr}$ for fixed large N, indicting that the Dicke model can be
solved easily in these two limits. More interestingly, just in the
critical regime, a dip structure is found in all curves, indicating
a trace of QPT at finite size system.

\begin{figure}[tbp]
\centering
\includegraphics[width=8cm]{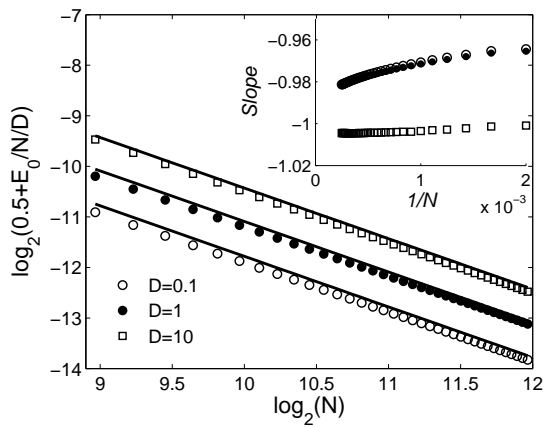}
\caption{Scaling of the ground-state energy $E_{0}/ND+0.5$ as a
function of $ N$ at the critical point for $D=0.1, 1$, and $10$. The
inset show the corresponding slope versus $1/N$. } \label{fig3}
\end{figure}

To study the universality of the QPT, we will compute the finite-size
scaling exponents for several observables.

\textsl{Ground-state energy}.--We calculate the ground-state energy
for different values of $D$. In the thermodynamical limit,
$E_0/ND=-0.5$. We are able to study the system up to $N=2^{12}$
atoms in a PC. To show the leading finite-size corrections, we plot
$E_0/ND+0.5$ versus $N$ for $D=0.1,1,10$ in a log-log scale in Fig.
3. The inset presents the slope as a function $1/N$. It is observed
that the slope of all these curves in the large $N$ regime give a
universal exponent $-1.0\pm 0.02$, from the above and below
respectively, consistent with that by a modified Holstein-Primakoff
approach\cite{vidal}.  The asymptotic behavior demonstrates that the
exponent by no means lies on the vicinity of $-4/3$ in the the
adiabatic regime reported in Ref. \cite{liberti}.

\begin{figure}[tbp]
\centering
\includegraphics[width=8cm]{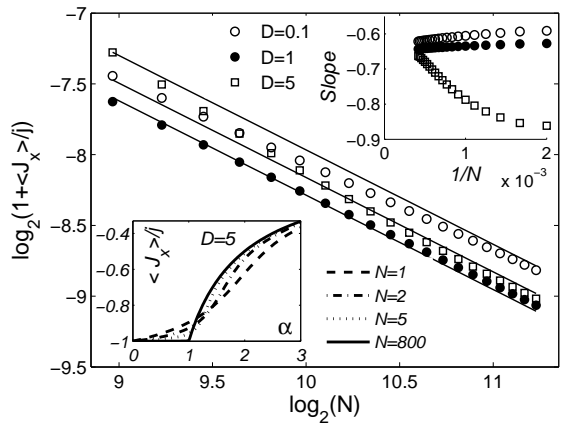}
\caption{Scaling of the ground-state Berry phase $1+<J_{x}>/N $ as a
function of $N$ at the critical point for $D=0.1, 1$, and $5$. The
upper inset show the corresponding slope as a function $1/N$, and
the lower one gives the Berry phase versus $\alpha$ for different
$N$ at $D=5$. } \label{fig4}
\end{figure}
\textsl{Berry phase}.-- To generate the Berry phase, we introduce a
time-dependent unitary transformation $U(T)=exp[-i\phi (t)J_x]$,
where $\phi (t)$ is changed adiabatically and slowly from $0$ to
$2\pi $. The ground-state Berry phase can be defined
as\cite{plastina}
\begin{equation}
\gamma =i\int_0^{2\pi }<\varphi |U^{+}(t)\frac d{d\phi }U(t)|\varphi >d\phi
=2\pi <\varphi |J_x|\varphi >  \label{BP}
\end{equation}
where $|\varphi >$ is just the time-independent ground-state
wavefunction, which can be obtained by the present DCS. In the
thermodynamical limit $<J_x>/N=-1$.

Fig. 4 shows the scaled ground-state Berry phase $1+<J_x>/N$ as a
function of $N$ for different values of $D$ in log-log scale. A
power law behavior exists in the large $N$. One can see from the
upper inset that the finite-size exponents extracted from all curves
tend to a converging value $(-0.67\pm 0.02)$. This result is in good
agreement with that in Ref. (\cite{plastina}) with a adiabatic
approximation. We find that the scaling exponent for the Berry phase
in Dicke model is also universal, not merely in the adiabatic limit.
The Berry phase tends to be discontinuous in the critical regime
with the increase of $ N$, shown in the lower inset.

\begin{figure}[tbp]
\centering
\includegraphics[width=8cm]{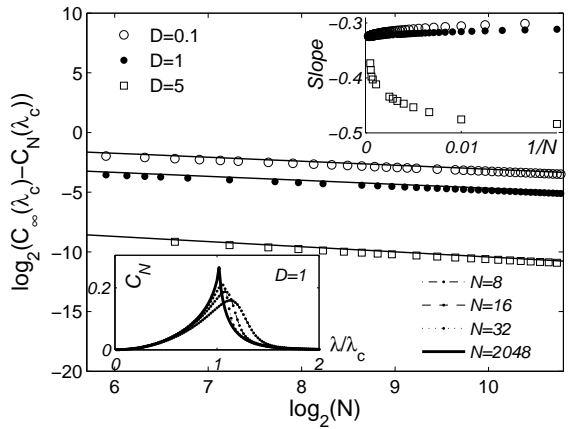}
\caption{Scaling of concurrence as a function of $N$ at the critical
point for $D=0.1, 1$, and $5$. The upper inset show the slope versus
$1/N$, the lower inset displays the scaled concurrence as a function of $%
\lambda/\lambda_c$.}
\label{fig5}
\end{figure}

\textsl{Pairwise entanglement}.--An important ground-state quantum
essential feature is the concurrence, which quantifies the
entanglement between two atoms in the atomic ensemble after tracing
out over bosons. The scaled concurrence (entanglement) in the
ground-state can be defined as $C_N=1-4<J_y^2>/N$ \cite
{vidal,wang}. In the thermodynamic limit $N\rightarrow \infty $, the
scaled concurrence at critical point $C_\infty (\lambda _c)$ can be
determined in terms of $D$ \cite{Lambert2}. For comparison, we can
calculate the quantity $ C_\infty (\lambda _c)-C_N(\lambda _c)$. In
Fig. 5, we present this quantity as a function of $N$ for different
values of $D$ in log-log scale. Derivative of these curves is
presented in the upper inset, and the exponent of concurrence is
estimated to be ${-0.33\pm 0.02}$. For large $D$, the exponent for
concurrence is only given reasonably in the asymptotic regime. From
the lower inset, we find that a cusp is formed with increasing $N$
in the critical regime, demonstrating the relic of the QPT in finite
size.

Lambert et al., performed a numerical DFS and give the exponent of
concurrence ${-0.25\pm 0.01}$ \cite{Lambert2} with $N\le 32$. Reslen
et al., derived a effective Hamiltonian, and found the exponent to
be ${-0.26\pm 0.02 }$ with $N\le 35$.\cite{reslen} Recently, Vidal
et al. using the diagonalizing a expanded Hamiltonian at order $1/N$
based on Holstein-Primakoff representation, and predicted
finite-size scaling exponent for concurrence $ 1/3$ \cite{vidal},
agreeing well with our results. Due to the huge system size we can
touch here, the previous controversy can be attributed to the small
system size \cite{Lambert2,reslen}.

In summary, we have proposed a numerically exact solution to the
Dicke model in the whole coupling regime for a huge system size up
to $N=2000 \thicksim 4000$, which is almost two orders of magnitude
higher than that reported in literature \cite{Lambert2,reslen}.
Several quantities related to QPT such as the ground-state energy,
the Berry phase, and the concurrence are then calculated exactly. It
is found that the scaling exponent for the ground-state energy is
different from that in a previous study\cite{liberti}. The previous
discrepancy for the exponent of concurrence is reconciled by the
present calculation in very large size systems. The precise estimate
of the scaling exponent for these quantities is  very significant to
clarify the universality of QPT. The methodology sketched here is
helpful to study a large number of spin(fermion)-boson coupling
systems.

This work was supported by National Natural Science Foundation of
China, PCSIRT (Grant No. IRT0754) in University in China,  and
National Basic Research Program of China (Grant No. 2009CB929104).

\end{document}